# Empirical Relationships among Lepton and Quark Masses

**Alan Breakstone**

*887 West Knickerbocker Drive, Sunnyvale, CA 94087*

**Abstract**

We derive empirical relationships among elementary fermion masses based on relatively simple exponential formulae involving quantum numbers for the electromagnetic and strong interactions, with a weak correction factor motivated by a simple linear combination of mass terms for the weak and electromagnetic interactions of charged leptons. This results in a prediction of $m_\tau = $ 1776.81234 (33) *MeV*, to be compared to the world average measurement of 1776.84 (17) *MeV*. We find a simple $2\pi$ relationship among neutrino mass ratios agrees well with present neutrino oscillation measurements. We extend the form of the mass ratios to the quark sector, giving qualitative agreement with quark masses.

## 1. Introduction

The masses of the elementary fermions in the Standard Model seem to have arbitrary values. There is a clear hierarchy in the sense that masses increase as one goes from the first to third generation. For a given generation the mass of the neutrino is much less than that of the charged lepton, which is, in turn, less than that of the associated quarks. In this work we propose a set of empirical relations for mass ratios among the generations as well as between the lepton and quark sectors.

We begin with a relationship among the generations of charged leptons, then propose a possible relation among the generations of neutrinos, then continue with relations among the generations of up-type and down-type quarks. Finally we propose an empirical relationship that connects the lepton masses with those of the quarks.

## 2. Charged Lepton Mass Ratios

Table 1 gives the Codata 2006 [1] compilation of charged lepton masses and mass ratios.

|   | Mass (*MeV*) |
|---|---|
| e | 0.510998910 (13) |
| μ | 105.6583668 (38) |
| τ | 1776.99 (29) |
|   | Mass ratio |
| μ/e | 206.7682823 (52) |
| τ/e | 3477.48 (57) |

Table 1: Codata 2006 compilation of charged lepton masses and mass ratios. The errors in the last two digits are given in parentheses.

In the early 1980's, Koide [2] found the following relation among these masses:

$$m_e + m_\mu + m_\tau = \frac{2}{3}(\sqrt{m_e} + \sqrt{m_\mu} + \sqrt{m_\tau})^2 \tag{1}$$

This formula fits the experimental data remarkably well.

In this paper we take a different empirical approach to the mass ratios of the charged leptons. We begin by noting that the natural logarithms of the mass ratios are very near to certain rational numbers.

$$\ln\left(\frac{m_\mu}{m_e}\right) = 5.3316 \approx \frac{16}{3} \tag{2}$$

$$\ln\left(\frac{m_\tau}{m_e}\right) = 8.154 \approx \frac{49}{6} \tag{3}$$

If we assign generation numbers to the charged leptons with $G = 1$ for the electron, $G = 2$ for the muon, and $G = 3$ for the tau, the relations (2) and (3) can be replaced with a single relation

$$\ln\left(\frac{m_x}{m_e}\right) \approx \frac{(3\Delta G + 1)^2}{3\Delta G} \tag{4}$$

Where $x$ is either $\mu$ or $\tau$ and $\Delta G$ is the difference in generation numbers.

The question arises as to the origin of the factor of 3 in equation (4). One possibility is that it is related to the electric charge of the charged leptons. As we shall see later, when investigating mass ratio relationships in the quark sector, it is convenient to define $Q = 3|q|$, where $q$ is the charge in elementary units. Then we rewrite equation (4) a bit more generally as

$$\ln\left(\frac{m_x}{m_e}\right) \approx \frac{(Q\Delta G + 1)^2}{Q\Delta G} \tag{5}$$

At this point the mass relation is just approximate. If we associate the mass in some sense with the interactions of elementary particles, then the charged leptons should have a contribution from the weak interaction as well as the electromagnetic interaction, which presumably dominates the mass. We postulate that the measured mass is a simple sum of weak and electromagnetic terms and that the electromagnetic term follows equation (5) exactly, i.e.,

$$m_e = m_e^{em} + m_e^{weak} \tag{6}$$

$$\ln\left(\frac{m_x^{em}}{m_e^{em}}\right) = \frac{(Q\Delta G + 1)^2}{Q\Delta G} \tag{7}$$

Equation (6) can be rewritten for each of the charged leptons as

$$m_e = m_e^{em}(1 + \varepsilon_1), \text{ where } \varepsilon_1 = \frac{m_e^{weak}}{m_e^{em}} \tag{8a}$$

$$m_\mu = m_\mu^{em}(1 + \varepsilon_2), \text{ where } \varepsilon_2 = \frac{m_\mu^{weak}}{m_\mu^{em}} \tag{8b}$$

$$m_\tau = m_\tau^{em}(1 + \varepsilon_3), \text{ where } \varepsilon_3 = \frac{m_\tau^{weak}}{m_\tau^{em}} \tag{8c}$$

Equation (7) then implies

$$m_\mu^{em} = m_e^{em} e^{16/3} = \left[\frac{m_e}{1+\varepsilon_1}\right] e^{16/3} = \frac{m_\mu}{1+\varepsilon_2} \tag{9a}$$

$$m_\tau^{em} = m_e^{em} e^{49/6} = \left[\frac{m_e}{1+\varepsilon_1}\right] e^{49/6} = \frac{m_\tau}{1+\varepsilon_3} \tag{9b}$$

Therefore

$$\varepsilon_2 = \frac{m_\mu}{m_e}(1+\varepsilon_1)e^{-16/3} - 1 \tag{10a}$$

$$\varepsilon_3 = \frac{m_\tau}{m_e}(1+\varepsilon_1)e^{-49/6} - 1 \tag{10b}$$

Plugging in the measured masses gives negative values for $\varepsilon_1$, $\varepsilon_2$, and $\varepsilon_3$. This implies that the measured mass is diminished by the weak contribution, rather than augmented by it, as might be suggested by a composite model of charged leptons. We continue the empirical approach by keeping the $\varepsilon$'s positive and incorporating the negative sign into equation (6), i.e.,

$$m_e = m_e^{em} - m_e^{weak} \tag{11}$$

and similarly for the other charged leptons. Equations (9a) and (9b) now become

$$m_\mu^{em} = m_e^{em} e^{16/3} = \left[\frac{m_e}{1-\varepsilon_1}\right] e^{16/3} = \frac{m_\mu}{1-\varepsilon_2} \qquad (12a)$$

$$m_\tau^{em} = m_e^{em} e^{49/6} = \left[\frac{m_e}{1-\varepsilon_1}\right] e^{49/6} = \frac{m_\tau}{1-\varepsilon_3} \qquad (12b)$$

yielding

$$\varepsilon_2 = 1 - \frac{m_\mu}{m_e}(1-\varepsilon_1)e^{-16/3} \qquad (13a)$$

$$\varepsilon_3 = 1 - \frac{m_\tau}{m_e}(1-\varepsilon_1)e^{-49/6} \qquad (13b)$$

Let us further suppose that the weak terms have a simple hierarchy as follows.

$$\frac{\varepsilon_2}{\varepsilon_1} = \frac{\varepsilon_3}{\varepsilon_2} = a \qquad (14)$$

Then

$$\varepsilon_2 = a\varepsilon_1 = 1 - \frac{m_\mu}{m_e}(1-\varepsilon_1)e^{-16/3} \qquad (15a)$$

$$\varepsilon_3 = a^2\varepsilon_1 = 1 - \frac{m_\tau}{m_e}(1-\varepsilon_1)e^{-49/6} \qquad (15b)$$

If we define

$$R_\mu \equiv \frac{m_\mu}{m_e}, \quad R_\tau \equiv \frac{m_\tau}{m_e}, \quad c_\mu \equiv e^{16/3}, \quad c_\tau \equiv e^{49/6} \qquad (16)$$

Then equations (15a) and (15b) can be solved for $\varepsilon_1$ giving

$$\varepsilon_1 = \frac{c_\mu - R_\mu}{ac_\mu - R_\mu} \qquad (17a)$$

$$\varepsilon_1 = \frac{c_\tau - R_\tau}{a^2 c_\tau - R_\tau} \qquad (17b)$$

Equations (17a) and (17b) can be combined to give a quadratic equation for $a$

$$a^2 c_\tau - a c_\mu \left( \frac{c_\tau - R_\tau}{c_\mu - R_\mu} \right) - R_\tau + R_\mu \left( \frac{c_\tau - R_\tau}{c_\mu - R_\mu} \right) = 0 \tag{18}$$

Putting in the Codata 2006 values for the charged lepton masses, the positive root solution gives $a = 6.2263$. The negative root solution gives the trivial $a = 1$ solution. The positive root value is rather close to $2\pi$. In the remainder of this paper we will set $a = 2\pi$ which will allow us to calculate $\varepsilon_1$ based on the measured electron and muon masses, then make a prediction for the tau mass.

Thus, we now have for the measured masses

$$\frac{m_\mu}{m_e} = \frac{m_\mu^{em}(1-\varepsilon_2)}{m_e^{em}(1-\varepsilon_1)} = \frac{(1-2\pi\varepsilon_1)}{(1-\varepsilon_1)} e^{16/3} \tag{19a}$$

$$\frac{m_\tau}{m_e} = \frac{m_\tau^{em}(1-\varepsilon_3)}{m_e^{em}(1-\varepsilon_1)} = \frac{(1-(2\pi)^2 \varepsilon_1)}{(1-\varepsilon_1)} e^{49/6} \tag{19b}$$

These can be re-written in a more general form

$$\frac{m_x}{m_e} = \frac{1-(2\pi)^{\Delta G} \varepsilon_1}{1-\varepsilon_1} e^{[(Q\Delta G+1)^2 / Q\Delta G]} \tag{20}$$

where $x = \mu$ or $\tau$.

Using the measured value of $m_\mu / m_e = 206.7682823\ (52)$ from the Codata 2006 compilation in equation (19a), we calculate $\varepsilon_1 = 0.0003279280\ (47)$. Using this value of $\varepsilon_1$ in equation (19b) then gives a prediction that $m_\tau = 1776.81234\ (47)$ *MeV*, which is remarkably close to current (2008) world average measured value [3] of 1776.84 (17) *MeV*. It is important to note that equations (19a), (19b), and (20) contain two relationships and only one free parameter, $\varepsilon_1$, which when determined from experimental data, results in an unambiguous and fully constrained prediction for a separate experimentally determinable quantity, the $\tau$ mass.

It is interesting to compare this prediction with the prediction from the Koide formula, equation (1), which gives $m_\tau = 1776.969\ (19)$ *MeV*. The difference between this prediction and that of equation (19b) is 157 *keV*, slightly below the measurement error, which is dominated by the BES (1996) [4], KEDR (2007) [5], and BELLE (2007) [6] measurements of the $\tau$ mass. The BESIII collaboration [7] plans to do a new $\tau$ mass measurement, hopefully with lower statistical and systematic errors than the KEDR and BELLE measurements. If they measure the mass to a precision of 50 *keV* or better, this measurement should unambiguously distinguish between these two predictions. Note that the current central value for the measurement is just 30 *keV* above the prediction of this paper and is 130 *keV* below Koide's prediction.

To summarize, although there were a number of somewhat arbitrary assumptions made in arriving at the mass relation given in equation (20), the end result agrees very well with the τ mass measurement. It predicts a slightly different mass from that of the Koide formula, allowing experimental differentiation of the differing predictions. As we shall see later in this paper the form of equation (20) suggests an extension to mass ratio relationships among the quarks.

### 3. Neutrino Mass Ratios

As is well known, the existence of neutrino oscillations requires neutrinos to have non-zero masses. Recent KamLAND measurements for reactor neutrino oscillations [8] give $\Delta m_{12}^2 = 7.9^{+0.6}_{-0.5} \times 10^{-5} eV^2$. Super-Kamiokande results for atmospheric neutrino oscillations [9] give $1.9 \times 10^{-3} < \Delta m_{23}^2 < 3.0 \pm 0.3 \times 10^{-3} eV^2$ at 90%CL. If we assume that the neutrino masses are associated with the weak interaction and that their masses are in the same ratio as for the weak contribution to the charged lepton masses, then we propose

$$m_3 = 2\pi m_2 \,, \quad m_2 = 2\pi m_1 \tag{21}$$

so that equation (20) can be re-written as

$$\frac{m_x}{m_e} = \frac{1 - \frac{m_i}{m_1}\varepsilon_1}{1 - \varepsilon_1} e^{\left[(Q\Delta G+1)^2 / Q\Delta G\right]} \tag{22}$$

where $i = 2$ for $x = \mu$ and $i = 3$ for $x = \tau$.

Combining equation (21) with the reactor neutrino oscillation data, one can calculate the mass of the first generation neutrino as follows.

$$\Delta m_{12}^2 \equiv m_2^2 - m_1^2 = 4\pi^2 m_1^2 - m_1^2 = (4\pi^2 - 1)m_1^2 \tag{23}$$

$$m_1 = \sqrt{\frac{\Delta m_{12}^2}{4\pi^2 - 1}} \tag{24}$$

Putting in the measurement gives $m_1$ = 0.00143 (5) eV. Equations (21) then give $m_2$ = 0.0090 (3) *eV* and $m_3$ = 0.057 (2) *eV*. Using these values for $m_2$ and $m_3$, we predict $\Delta m_{23}^2 = 3.1 \pm 0.2 \times 10^{-3} eV^2$, which is a bit high compared to the measurement. One can get better agreement with both measurements by fitting the neutrino masses using both the reactor and atmospheric neutrino oscillation measurements. This gives $m_1$ = 0.00139 (6) *eV*, $m_2$ = 0.0087 (4) *eV*, and $m_3$ = 0.055 (2) *eV*, which result in a value of $\Delta m_{12}^2$ about 0.8 standard deviations below the central value of the measurement and a value of $\Delta m_{23}^2$ about 1.2 standard deviations above the central value of the measurement. Clearly the data are not precise enough to rule out the mass relations given in equation (21), although these relations are only weakly motivated.

## 4. Quark Mass Ratios

Since quarks are confined within hadrons, quark masses are not as well determined as lepton masses. Additionally quark masses depend upon the theoretical framework used to define them [10]. Nonetheless we can extend the methods used in the previous sections to find relationships between the generations of quarks. The quark mass ratios will have a factor for the electromagnetic interaction and another factor for the strong interaction. Additionally they will have a correction for the weak interaction. For the electromagnetic interaction, we assume the form of equation (7). For the up-type quarks, $Q = 2$, so

$$\frac{m_c^{em}}{m_u^{em}} = e^{9/2} \; , \; \frac{m_t^{em}}{m_u^{em}} = e^{25/4} \tag{25}$$

Similarly, for the down-type quarks, $Q = 1$, so

$$\frac{m_s^{em}}{m_d^{em}} = e^4 \; , \; \frac{m_b^{em}}{m_d^{em}} = e^{9/2} \tag{26}$$

We assume that the strong interaction factor multiplies the electromagnetic factor, that the form of the strong factor is exponential, and that there is a weak factor which is the same as for the charged leptons, i.e.

$$\frac{m_x}{m_y} = \frac{m_x^{em}}{m_y^{em}} \cdot \frac{m_x^{strong}}{m_y^{strong}} \cdot F_{weak} \tag{27}$$

$$\frac{m_x^{strong}}{m_y^{strong}} = e^{a\Delta G + b} \tag{28}$$

$$F_{weak} = \frac{1 - (2\pi)^{\Delta G} \varepsilon_1}{1 - \varepsilon_1} \tag{29}$$

After a little experimentation, we find that $a = 3$, $b = -2$ works reasonably well for up-type quarks, while $a = 3$, $b = -4$ works for down-type quarks. If we rewrite the form of the exponent as

$$3\Delta G - 2 = 3(\Delta G - 1) + 1 \tag{30a}$$
$$3\Delta G - 4 = 3(\Delta G - 1) - 1 \tag{30b}$$

we can identify the first term with the baryon number of the quarks, $B$, and the second term with isospin component $I_z$, and write the strong factor as

$$\frac{m_x^{strong}}{m_y^{strong}} = e^{3\beta(\Delta G-1)} e^{2I_z} \tag{31}$$

where $\beta = 3B = 3 (1/3) = 1$ and $I_z = \pm \frac{1}{2}$.

Putting this all together for the quark mass ratios we get

$$\frac{m_x}{m_{u,d}} = \frac{1-(2\pi)^{\Delta G} \varepsilon_1}{1-\varepsilon_1} e^{\frac{(Q\Delta G+1)^2}{Q\Delta G}} e^{3\beta(\Delta G-1)} e^{2I_z} \tag{32}$$

where $x = t$ and $c$ for the up-type quarks and $x = b$ and $s$ for the down-type quarks.

Since the masses of the heavier quarks are better determined than those of the light quarks, it makes the most sense to use those to obtain the numerical values for the quark masses which are given in Table 2.

|   | PDG Compilation (*MeV*) | Calculated Mass (*MeV*) |
|---|---|---|
| u | 1.5 to 5 | $6.2 \pm 0.1$ |
| d | 5 to 9 | $6.5 \pm 0.5$ |
| c | 1000 to 1400 | $1511 \pm 36$ |
| s | 80 to 155 | $130 \pm 9$ |
| t | $172700 \pm 1700 \pm 2400$ | |
| b | $4260 \pm 150 \pm 150$ | |

Table 2: Masses of quarks. The top mass is from Fermilab measurements [11], the rest of the masses in the second column come from the PDG 2004 compilation [10]. Masses in the third column are calculated using equation 31. Errors for the calculated masses use a sum of statistical and systematic errors from the measurements of the top or bottom mass.

As one can see equation (32) gives reasonable agreement with the PDG compilation of quark masses, although the up and charm quark masses are somewhat high.

### 5. Mass Ratios within Generations

Guided by the form of equation (32), we propose a relationship of masses within a generation of the form

$$\frac{m_x}{m_y} \approx e^{a_G \Delta Q + b_G \Delta I_z + c_G \Delta \beta} \tag{33}$$

where $\Delta Q = 3(|q_x| - |q_y|)$, $\Delta I_z = I_{zx} - I_{zy}$, and $\Delta \beta = 3(B_x - B_y)$ (34)

This gives a set of nine relationships as follows

$$m_t \approx m_\tau e^{-a_3 + c_3} \tag{35a}$$

$$m_b \approx m_\tau e^{-2a_3 - b_3 + c_3} \tag{35b}$$

$$m_3 \approx m_\tau e^{-3a_3 - b_3} \tag{35c}$$

$$m_c \approx m_\mu e^{-a_2 + c_2} \tag{35d}$$

$$m_s \approx m_\mu e^{-2a_2 - b_2 + c_2} \tag{35e}$$

$$m_2 \approx m_\mu e^{-3a_2 - b_2} \tag{35f}$$

$$m_u \approx m_e e^{-a_1 + c_1} \tag{35g}$$

$$m_d \approx m_e e^{-2a_1 - b_1 + c_1} \tag{35h}$$

$$m_1 \approx m_e e^{-3a_1 - b_1} \tag{35i}$$

Consistency with the intergenerational formulas, equations (20), (21), and (33) provide six constraint equations among the constants $a_i$, $b_i$, and $c_i$ in the exponents as well as two constraints involving the weak correction term. The intergenerational formulas can be written as

$$m_t = m_u F_2 e^{41/4} \tag{36a}$$

$$m_c = m_u F_1 e^{11/2} \tag{36b}$$

$$m_b = m_d F_2 e^{13/2} \tag{36c}$$

$$m_s = m_d F_1 e^3 \tag{36d}$$

$$m_\tau = m_e F_2 e^{49/6} \tag{36e}$$

$$m_\mu = m_e F_1 e^{16/3} \tag{36f}$$

$$m_3 = m_1 (2\pi)^2 \tag{36g}$$

$$m_2 = m_1 (2\pi) \tag{36h}$$

where we define $F_{\Delta G} \equiv \dfrac{1 - (2\pi)^{\Delta G} \varepsilon_1}{1 - \varepsilon_1}$ (36i)

Numerically, $F_1 = 0.998266935$ (35) and $F_2 = 0.98737764$ (26).

The six constraint equations among the constants $a_i$, $b_i$, and $c_i$ in the exponents are

$$-a_1 + c_1 = -a_3 + c_3 - \frac{25}{12} \tag{37a}$$

$$-2a_1 - b_1 + c_1 = -2a_3 - b_3 + c_3 + \frac{10}{6} \tag{37b}$$

$$-a_1 + c_1 = -a_2 + c_2 - \frac{1}{6} \tag{37c}$$

$$-2a_1 - b_1 + c_1 = -2a_2 - b_2 + c_2 + \frac{7}{3} \tag{37d}$$

$$-3a_1 - b_1 = -3a_3 - b_3 + \frac{49}{6} \tag{37e}$$

$$-3a_1 - b_1 = -3a_2 - b_2 + \frac{16}{3} \tag{37f}$$

Using the third generation mass ratios and equations (35a), (35b), and (35c), we can get numerical values for $a_3$, $b_3$, and $c_3$.

$$\frac{m_t}{m_b} = \frac{172700}{4260} = 40.5 \approx e^{15/4} \Rightarrow a_3 + b_3 = \frac{15}{4} \tag{38a}$$

$$\frac{m_t}{m_\tau} = \frac{172700}{1776.99} = 97.2 \approx e^{9/2} \Rightarrow -a_3 + c_3 = \frac{9}{2} \tag{38b}$$

$$\frac{m_\tau}{m_3} = \frac{1776.99}{5.6 \times 10^{-8}} = 3.2 \times 10^{10} \approx e^{97/4} \Rightarrow 3a_3 + b_3 = \frac{97}{4} \tag{38c}$$

Equations (38a-c) give $a_3 = \frac{41}{4}$, $b_3 = -\frac{13}{2}$, and $c_3 = \frac{59}{4}$.

The constraint equations (37a-f) then give the rest of the values for $a_i$, $b_i$, and $c_i$ given in Table 3.

| Generation | a | b | c |
|---|---|---|---|
| 1 | 193/24 | -193/24 | 251/24 |
| 2 | 227/24 | -167/24 | 289/24 |
| 3 | 246/24 | -156/24 | 354/24 |

Table 3: Values of constants in the exponents for mass ratios within the generations.

The equations (35a-i) are only approximate. We can make them exact by introducing two constants, one for the ratio of quark to charged lepton masses, the other for the ratio of neutrino to charged lepton masses. To be precise, we define them using the now exact equations

$$m_t = m_\tau F_q e^{-a_3 + c_3} \tag{39a}$$

$$m_3 = m_\tau F_\nu e^{-3a_3 - b_3} \tag{39b}$$

Using the experimentally measured top mass and the $m_3$ mass from section 3 we have numerically $F_q = 1.08 \pm 0.02$ and $F_v = 1.05 \pm 0.04$. Putting in the factors $F_1$ and $F_2$ as well, we can write exact expressions for equations (35a-i).

$$m_t = m_\tau F_q e^{-a_3 + c_3} \tag{40a}$$

$$m_b = m_\tau F_q e^{-2a_3 - b_3 + c_3} \tag{40b}$$

$$m_3 = m_\tau F_v e^{-3a_3 - b_3} \tag{40c}$$

$$m_c = m_\mu F_q e^{-a_2 + c_2} \tag{40d}$$

$$m_s = m_\mu F_q e^{-2a_2 - b_2 + c_2} \tag{40e}$$

$$m_2 = m_\mu F_v \left( \frac{F_2}{2\pi F_1} \right) e^{-3a_2 - b_2} \tag{40f}$$

$$m_u = m_e F_q e^{-a_1 + c_1} \tag{40g}$$

$$m_d = m_e F_q e^{-2a_1 - b_1 + c_1} \tag{40h}$$

$$m_1 = m_e F_v \left( \frac{F_2}{(2\pi)^2} \right) e^{-3a_1 - b_1} \tag{40i}$$

Using equations (40a-i) and (20), the constants $\varepsilon_1$, $F_q$, $F_v$, $a_3$, $b_3$, $c_3$, plus the measured electron mass, one can calculate the rest of the masses. The numerical results of this calculation are given in Table 4.

|   | Measured Mass (*MeV*) | Calculated Mass (*MeV*) |
|---|---|---|
| e | 0.510998910 (13) | |
| μ | 105.6583668 (38) | 105.6583669 (38) |
| τ | 1776.84 (17) | 1776.81234 (33) |
| $v_1$ | $1.39(6) \times 10^{-9}$ | $1.39(5) \times 10^{-9}$ |
| $v_2$ | $8.7(4) \times 10^{-9}$ | $8.8(3) \times 10^{-9}$ |
| $v_3$ | $5.5(2) \times 10^{-8}$ | $5.5(2) \times 10^{-8}$ |
| u | 1.5 to 5 | 6.18 (10) |
| c | 1000 to 1400 | 1511 (25) |
| t | $172700 \pm 1700 \pm 2400$ | 172700 (2900) |
| d | 5 to 9 | 6.18 (10) |
| s | 80 to 155 | 124.0 (2.1) |
| b | $4260 \pm 150 \pm 150$ | 4062 (96) |

Table 4: Measured and calculated masses. The measured masses come from Codata 2006 [1] and PDG 2008 [3] for the charged leptons, neutrino oscillation measurements [5,6] plus the assumptions in section 3 for the neutrinos, the PDG 2004 compilation [10] for the quarks, except for the top quark which is from recent Fermilab results [11]. The calculations use equations (20) for the charged leptons and (40a-i) for the others.

The values from the intragenerational formulas, equations (40a-i), give slightly different numbers than those derived from the intergenerational formulas because of the different assumptions that go into the two cases. For the intragenerational formulas, the constants $F_q$ and $F_v$ are defined to make the top and $\nu_3$ masses agree with the experimentally measured ones, which then skew some of the other masses. As with the intergenerational formulas, the calculated up and charm quark masses are somewhat large.

## 6. Summary and Conclusions

As noted throughout this paper, these calculations are empirically derived. There is, unfortunately, no theoretical understanding of the form of the equations. There have been a number of attempts to derive deeper theoretical understanding of the Koide mass formula for charged leptons [12,13] and to extend it to the quark sector [14]. Additionally there have been string theory motivated results for both charged leptons and neutrinos [15].

As with the Koide formula, this work makes a definite and accurate prediction of the $\tau$ mass. It is important to note that since this prediction was originally made in 2006, new $\tau$ mass measurements have moved the world average in the direction of this prediction. If the BESIII collaboration or other experiments to measure the $\tau$ mass measure it with sufficient accuracy, one will be able to further determine whether or not this approach to mass formulas has any merit. For the neutrinos, future oscillation experiments will provide accurate values for $m_{12}$ and $m_{23}$, which will then allow us to see if the simple $2\pi$ relationship given in section 3 is valid for neutrino mass ratios.

For the quark sector, these calculations give qualitative agreement with measured values, but are not completely satisfying. There is clearly additional work needed to understand whether or not this approach has validity.

Most importantly a theory needs to be developed that gives the form of the mass relationships and predicts the values of the six experimentally determined constants. The simple form of the mass ratio relationships and their linkage with the weak, electromagnetic, and strong interactions, if correct, are remarkable. Hopefully this work is the first step towards a deeper understanding of the values of elementary fermion masses.